\documentclass[aps,pre,amsmath,amsfonts,floatfix,twocolumn]{revtex4}

\usepackage{graphicx}
\let\a=\alpha \let\b=\beta  \let\g=\gamma

\let\s=\sigma

  \let\r=\rho 

\def\ie{{i.e. }}\def\eg{{e.g. }}

\def\to{\rightarrow}

\newcommand{\beq}{\begin{equation}}
\newcommand{\eeq}{\end{equation}}

\begin{document}

\title{
Comment to ``Packing Hyperspheres in High-Dimensional Euclidean Space''
}

\author{Francesco Zamponi\footnote{francesco.zamponi@phys.uniroma1.it}
}

\affiliation{Laboratoire de Physique Th\'eorique de l'\'Ecole Normale Sup\'erieure,
24 Rue Lhomond, 75231 Paris Cedex 05, France}

\date{\today}


\maketitle

In a recent paper \cite{SDST06} Skoge et al. presented new 
accurate numerical data on amorphous packings of hard spheres in space dimension $d > 3$.
It was shown (see \eg Fig.~4 in \cite{SDST06}) that on compressing the low
density liquid at a constant rate $\g$, the pressure of the system follows the 
equilibrium pressure of the liquid up to some density
(often called {\it glass transition density}) above which the pressure starts
to increase faster than in equilibrium, and diverges on approaching
a value of density $\phi_J(\g)$, which is called {\it jamming density}.
Crystallization seems to be strongly suppressed by kinetic effects in
dimension $d>3$ and can then be neglected in the following discussion.
Values of $\phi_J(\gamma)$ have been accurately 
measured in \cite{SDST06} as a function of $\gamma$. On the contrary,
$\phi_g(\gamma)$ is
not precisely determined as long as $\g > 0$: the glass transition is smeared
and happens in a crossover region $[\phi^-_g(\g),\phi^+_g(\g)]$. However, the amplitude
of the crossover interval seems to decrease for $\g \to 0$ (see again Fig.4 in
\cite{SDST06}, and see \cite{MGS06} for a recent theoretical discussion of
these effects).

Recently, a theory of the glass transition of hard spheres, that
can be applied in any space dimension $d$, was 
developed~\cite{PZ05,Za06}.
The basic idea is that the phenomenology observed at finite $\g$ is due to an
underlying {\it thermodynamic glass transition} \cite{KTW87b,MP99b,LW06}: 
\ie that in the limit 
$\g \to 0$, $\phi_g^\pm(\g) \to \phi_K$, the Kauzmann density, at which an
equilibrium phase transition to a glass phase happens. This is an idealized
picture that neglects metastability effects due to the presence of the
crystal: the consequences of this approximation might be important but cannot
be discussed here,
see \cite{Za06} for a detailed discussion. However, as it seems that
crystallization is negligible in $d>3$, a comparison between the theory and
the data of \cite{SDST06} is possible. In the equilibrium glass phase,
the pressure increases faster than in the liquid phase (\ie the 
compressibility is smaller), and diverges at a
value of density $\phi_{RCP} = \lim_{\g \to 0} \phi_J(\g)$ that we call
{\it random close packing (RCP) density}. This is the definition of
random close packing that can be given within our theory. However it is
affected by the metastability effects related to the presence of the crystal:
see \cite{To95,Za06} for a discussion. In particular, in \cite{To95} the
notion of random close packing density has been criticized and an alternative
notion of {\it maximally random jammed (MRJ) packings} has been proposed,
see below.

Our theory is based on standard liquid theory and on the replica trick: it
takes as input the equation of state of the liquid phase (in practice one has
to choose an expression that describes well the liquid at low density and
extrapolate to higher density), and gives as an output:
{\it i)} the Kauzmann density $\phi_K$; {\it ii)} the random close packing
density $\phi_{RCP}$; {\it iii)} the equation of state in the glass phase;
{\it iv)} some properties of the pair correlation function in the glass phase,
\eg its shape close to contact;
{\it v)} the equation of state of the metastable glass states that are reached
for $\g > 0$ and their contribution to the entropy 
(the {\it configurational entropy} or {\it complexity}); see \cite{PZ05}
for all the technical details.
One of the most interesting predictions of the theory is that 
$2^d \phi_{RCP} \sim d \log d$ for large $d$. As far as I know, 
this scaling has been proposed in \cite{PZ05}
for the first time. The aim of this paper is to show that the results of
\cite{SDST06} are fully compatible with this prediction.

\begin{table}[t]
\begin{tabular}{c|c|c|c|c}
\hline
$d$ & $\phi_K$ (theory) & $\phi_{RCP}$ (theory) & $\phi_{MRJ}$ \cite{SDST06} &
$\phi_{RCP}$ (extr.) \\
\hline
3 &  0.6175 & 0.6836 & 0.64 & --- \\
4 &  0.4319 & 0.4869 & 0.46 & 0.473 \\
5 &  0.2894 & 0.3307 & 0.31 & --- \\
6 &  0.1883 & 0.2182 & 0.20 & --- \\
7 &  0.1194 & 0.1402 & --- & --- \\
8 &  0.0739 & 0.0877 & --- & --- \\
\hline
\end{tabular}
\caption{
Values of $\phi_K$ and $\phi_{RCP}$ from the theory (only values
for $d\leq 8$ are reported for brevity, 
values for $d\geq 8$ are in Fig.~\ref{fig:scaling}) 
compared with the available measured
values of $\phi_{MRJ}$ \cite{SDST06}. The last column gives the value of
$\phi_{RCP}$ extrapolated from the data of \cite{SDST06} (see text).
}
\label{tab:I}
\end{table}

\begin{figure}
\includegraphics[width=8.5cm]{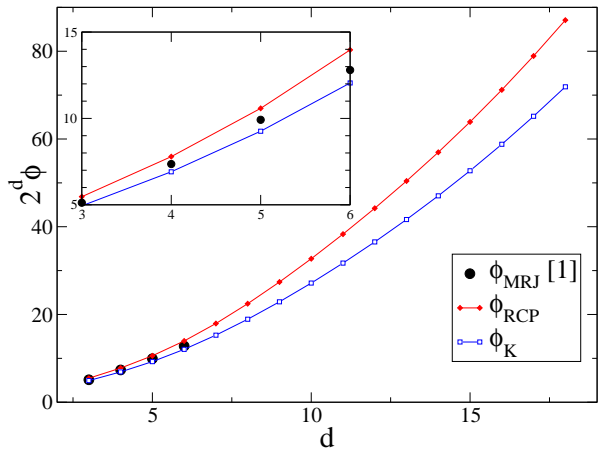}
\caption{Plot of $\phi_K$ (open squares, obtained solving Eq.~(\ref{phiKdef})), $\phi_{RCP}$ (full diamonds,
Eq.~(\ref{phiRCPdef})), and $\phi_{MRJ}$ (full circles, numerical
  estimate of \cite{SDST06}) as a function of the
  dimension. Both $\phi_K$ and $\phi_{RCP}$ scale as $2^d \phi \sim d \log d$
  for large $d$, while their distance scales as $2^d [\phi_{RCP}-\phi_K] \sim d$.
In the inset the same plot for $3\leq d \leq 6$ (compare with Fig.6 in
\cite{SDST06}).}
\vskip-10pt
\label{fig:scaling}
\end{figure}

To this aim, an expression for the equation of state of the liquid
in $d>3$ is needed as input to the theory. The simplest choice is a generalization of
the celebrated Carnahan-Starling equation of state to $d>3$ \cite{SMS89}:
\beq\begin{split}
&Y(\phi) = \frac{1-\a\phi}{(1-\phi)^d} \ , \\
&\a = d - 2^{d-1} (B_3/b^2) \ ,
\end{split}
\eeq
where $Y(\phi)=g(\s^+)$ is the value of the radial distribution function at contact,
and $b$ and $B_3$ are the second and third virial coefficients, whose exact
expression is known \cite{SMS89}. The entropy of the liquid $S(\phi)$ is obtained by
integrating the exact expression
\beq
\phi \frac{dS}{d\phi} = - \frac{\b P}{\r} = -1-b\r Y(\phi) \ .
\eeq
Given $S(\phi)$, the random close packing density is the solution of
\beq\label{phiRCPdef}
S(\phi) - d \log \left[ \frac{\sqrt{8}}{2^d \phi Y(\phi)} \right] + \frac{d}2
= 0 \ ,
\eeq
while the Kauzmann density is the solution of
\beq\label{phiKdef}
S(\phi) - d \log \left[ \frac{\sqrt{2\pi}}{2^d Q_0 \phi Y(\phi)} \right] =0 \ ,
\eeq
with $Q_0 = 0.638\ldots$ \cite{PZ05}. These equations can be easy solved
numerically to get the values of
$\phi_{RCP}$ and $\phi_K$ {\it for any given value of} $d$. The results are reported in
table~\ref{tab:I} for $d \leq 8$, and compared with $\phi_{MRJ}$ as reported in
\cite{SDST06}. The latter quantity is the density of the 
{\it maximally random} (according to some measure of ``order'') 
{\it collectively jammed packings} 
of the system, see \cite{To95} for the precise
definition; it is estimated in \cite{SDST06} by the jamming
density $\phi_J(\g)$ for finite but small $\g$ (see sec.IV in \cite{SDST06}
for a detailed discussion). As $\phi_J(\g)$ is expected to increase on
decreasing $\g$ and $\phi_{RCP} = \lim_{\g\to 0} \phi_J(\g)$, 
it follows that $\phi_{MRJ}$,
as estimated in \cite{SDST06},
should be strictly lower than $\phi_{RCP}$, but close to it,
consistently with the data in table \ref{tab:I}.
A plot of $\phi_K$ and $\phi_{RCP}$ for $d$ up to $19$ is reported 
in Fig.~\ref{fig:scaling}. Note that it has been suggested in \cite{SDST06} 
that $2^d \phi_{MRJ} \sim c_1 + c_2 d$; this scaling is not in contradiction with
$2^d \phi_K, 2^d \phi_{RCP} \sim d \log d$ because $\phi_{MRJ}$ does not
need to be bigger than $\phi_K$, even if this seems to be the case for $d \leq 6$.

The very nice data for $d=4$ reported in Fig.~4 of \cite{SDST06} allow for a more
precise comparison of the numerical and theoretical results: the value of
$\phi_J(\g)$ is reported for 5 different values of
$\g=10^{-3},10^{-4},10^{-5},10^{-6},10^{-7}$. A
standard procedure to extrapolate to $\g \to 0$ is to fit the data to a
Vogel-Fulcher law:
\beq\nonumber
\g(\phi_J) = \g_0 10^{-\frac{D}{\phi_{RCP}-\phi_J}} \Leftrightarrow
\phi_J(\g) = \phi_{RCP} + \frac{D}{\log_{10}[\g/\g_0]} \ .
\eeq
Such extrapolations are often ambiguous; however the fit is good and gives
$\phi_{RCP} = 0.473$, $D=0.03$, $\g_0 = 0.45$. The final result for
$\phi_{RCP}$ differs from $\sim 10\%$ from the theoretical value, see table~\ref{tab:I}.
This is a very good result given the ambiguities that are present both in the
numerical data (numerical error and extrapolation) and in the theory
(the choice of a particular expression for the equation of state of the liquid
that is not exact, see \cite{eqstate} for recent contributions).
Note that a similar extrapolation is not possible in $d=3$ due to
crystallization, and for $d>4$ due to lack of numerical data. 
Hopefully new data for $d>4$ will allow for a similar
comparison also in this case. Note also that the value of $\phi_K \sim 0.43$
we obtain in $d=4$ seems to agree very well with the extrapolation of
$\phi_g(\g)$ (defined roughly as the point where the curves leave the liquid
equation of state) to $\g=0$ in Fig.~4 of~\cite{SDST06}.

A more accurate comparison of the theory with the numerical data is
possible: for instance, the theory gives a prediction for the glass equation
of state that is close to the measured pressure in the glass branch for
very small $\g$, \eg $\g=10^{-7}$ in Fig.~4 of~\cite{SDST06};
it also predicts that the amorphous packings
are isostatic, \ie the average number of contacts per sphere is $z=2d$, 
in any dimension $d$, and this seems to be confirmed by the numerical data.
Other properties of the packings such as the shape of the correlation function
$g(r)$ close to contact are predicted to be independent of the dimension.
Unfortunately we are still not able to give a prediction for the splitting of
the second peak of $g(r)$, which seems to be strongly suppressed in $d>3$. We
hope that future work will address this and many other open questions 
\cite{Za06}. The data reported in \cite{SDST06} provide a very good test of the
theories aiming to understand the behavior of hard sphere in large space
dimension \cite{FP99,PS00,PZ05,Tordlarge} and strongly call for further developments.

\noindent
{\it Acknowledgments} - This paper is based on a collaboration with G.~Parisi:
I wish to thank him for his continuous support.
I wish to thank A.~Donev, S.~Torquato, and F.~H.~Stillinger for comments and
many useful discussions.
The work is supported by the EU Research Training Network STIPCO
(HPRN-CT-2002-00319).

\end{document}